\documentclass[reprint,amsmath,amssymb,aps,prb]{revtex4-1}

\pdfoutput=1

\usepackage{hyperref,url}
\usepackage{amsmath}
\usepackage{amsfonts}
\usepackage[capitalise]{cleveref}
\usepackage{placeins}
\hypersetup{breaklinks,colorlinks=true,linkcolor=blue,citecolor=blue,urlcolor=blue,filecolor=blue}
\usepackage{graphicx}
\usepackage{dcolumn}
\usepackage{subdepth}
\usepackage{tablefootnote}
\usepackage{multirow}
\usepackage{makecell}
\usepackage{booktabs}
\usepackage[normalem]{ulem}

\newcommand\mat\mathbf

\newcommand{\insertnew}[1]{{\textcolor{black} {#1}}}
\newcommand{\revrep}[2]{{\textcolor{red}{}}{\textcolor{black}{#2}}}

\begin{document}

\author{Joonho Lee}
\email{linusjoonho@gmail.com}
\affiliation{College of Chemistry, University of California, Berkeley, California 94720, USA.}

\author{Fionn D. Malone}
\email{malone14@llnl.gov}
\affiliation{Quantum Simulations Group, Lawrence Livermore National Laboratory, 7000 East Avenue, Livermore, CA, 94551 USA.}

\author{Miguel A. Morales}
\email{moralessilva2@llnl.gov}
\affiliation{Quantum Simulations Group, Lawrence Livermore National Laboratory, 7000 East Avenue, Livermore, CA, 94551 USA.}

\title{An Auxiliary-Field Quantum Monte Carlo Perspective on the Ground State of the Dense Uniform Electron Gas: An
Investigation with Hartree-Fock Trial Wavefunctions}
\begin{abstract}
We assess the utility of \revrep{spin-restricted Hartree-Fock (RHF)}{Hartree-Fock (HF)} trial wavefunctions in performing 
phaseless auxiliary-field quantum Monte Carlo (ph-AFQMC) on the uniform electron gas (UEG) model.
\revrep{This particular combination, RHF+ph-AFQMC}{The combination of ph-AFQMC with spin-restricted HF (RHF+ph-AFQMC)},
was found to be highly accurate and efficient for systems containing up to 114 electrons in 2109 orbitals, particularly for $r_s \le 2.0$.
Compared to spin-restricted coupled-cluster (RCC) methods, we found that RHF+ph-AFQMC performs better than CC with singles, doubles, and triples (RCCSDT) and similarly to or slightly worse than CC with singles, doubles, triples, and quadruples (RCCSDTQ) for $r_s \le 3.0$ in the 14-electron UEG model.
With the 54-electron, we found RHF+ph-AFQMC to be nearly exact for $r_s\le2.0$ and pointed out potential biases in existing benchmarks. Encouraged by these, we performed RHF+ph-AFQMC on the 114-electron UEG model for $r_s\le 2.0$ and provided new benchmark data for future method development. 
\revrep{As the UEG models with $r_s = 5.0$ remain to be challenging for RHF+ph-AFQMC, we emphasize the need for a better trial wavefunction for ph-AFQMC in simulating strongly correlated systems.}
{We found that the UEG models with $r_s=5.0$ remain to be challenging for RHF+ph-AFQMC.
Employing non-orthogonal configuration expansions or unrestricted HF states as trial wavefunctions was
also found to be ineffective in the case of the 14-electron UEG model with $r_s=5.0$. We emphasize the need for a better trial wavefunction for ph-AFQMC in simulating strongly correlated systems.} With the 54-electron and 114-electron UEG models, we stress the potential utility of RHF+ph-AFQMC  for simulating {\it dense} solids.
\end{abstract}
\maketitle
%\section{Fionn's To-Do List}
%\begin{enumerate}
%\item Check the section backpropagation if everything is correct.
%In particular, the path restoration weight update part needs to be reviewed. I used the hybrid energy form.
%\item Free-projection RDMs seem necessary for benchmarking rs = 5 in some small basis set.
%Or we need FCIQMC RDMs..
%\item Obtain raw data for 54 electrons FCIQMC from J. Shepherds.
%\item Write some things about UEG whenever necessary.
%\end{enumerate}
\section{Introduction \label{sec:intro}}
Describing electron correlation in a scalable way that can handle hundreds of electrons is a grand challenge in quantum chemistry and condensed matter physics.
State-of-the-art methods include coupled-cluster (CC) methods\citep{huntington2010pccsd,bartlett_rmp,small2012fusion,Kats2013,small2014coupled,lee2017coupled,lee2018open}, density matrix renormalization group (DMRG) methods\citep{white_dmrg_chem,chan_dmrg_chem}, and quantum Monte Carlo (QMC) approaches\citep{foulkes_rmp,zhang_phaseless,booth_fciqmc}. Each method exhibits different weaknesses and strengths and therefore they have been applied to solve a different class of problems in chemistry and condensed matter physics.

In this work, we will focus on a projector QMC method, namely, the auxiliary-field QMC (AFQMC) approach\citep{zhang_cpmc,zhang_phaseless}. 
Projector Monte Carlo methods, while formally exact, typically impose a constraint in the imaginary time propagation in order to overcome the fermion sign problem and achieve a polynomial scaling algorithm.
Both, diffusion Monte Carlo (DMC)\citep{foulkes_rmp} and AFQMC enforce this constraint using a trial wavefunction, which can in principle be systematically improved towards the exact result.  
These constraints lead to the phaseless-AFQMC (ph-AFQMC) and fixed-node (FN-DMC) algorithms both of which scale like $\mathcal{O}(N^3)-\mathcal{O}(N^4)$ with the number of electrons $N$.

Although the formalism of DMC and AFQMC are very similar, there are some key differences between the two.
First, AFQMC works in the second-quantized framework common to most quantum chemical methods, and introduces a finite basis set. Therefore, AFQMC energies need to be extrapolated to the complete basis set (CBS) limit in order to compare directly with experiments.
This is in contrast with DMC which works in real space and directly in the CBS limit.
Second, incorporating widely used Jastrow factors (JFs) into AFQMC is quite challenging. 
JFs are economical ways to incorporate residual electron correlation by enforcing cusp conditions either between electrons and nuclei or among electrons.
Lastly, unlike FN-DMC, ph-AFQMC is not variational\citep{carlson_no_var}.

%While these drawbacks can be somewhat disappointing, there are other features of AFQMC which make it quite attractive for simulating molecules and more generally {\it ab-initio} solids.
%This unique feature of AFQMC has already drawn some interests in quantum chemistry since it is most natural to perform all-electron calculations on systems with light elements[].
Despite these issues, AFQMC offers a number of promising advantages precisely because it works directly in an orbital-based basis.
In particular, all-electron, frozen core and non-local pseudopotential calculations can be performed with no additional approximations.
Furthermore, as most quantum chemistry methods are performed with a finite basis set, many tricks used in quantum chemistry can be used to improve AFQMC as well.
For instance, tensor hyper contraction approaches\citep{thc_1,thc_2,thc_3,lu_isdf,isdf_lin_1,isdf_lin_2} have recently been employed to reduce the memory requirement of AFQMC\citep{malone_isdf,motta_thc}.
Employing explicitly correlated basis functions (similar in spirit to JFs for DMC) should also be possible to reduce the basis set incompleteness error of AFQMC\citep{tew_expl_corr}.
In addition to this, computing properties other than the total energy, which has historically been a challenge for projector QMC methods, can be more straightforwardly achieved in AFQMC\citep{motta_bp}. Recent examples include one- and two-particle reduced density matrices\citep{motta_bp}, imaginary time correlation functions\citep{motta_itcf_1,motta_itcf_2,vitali_itcf} as well as forces\citep{motta_forces}.

%Since AFQMC is relatively unexplored in quantum chemistry, the scope of this method is in general not well understood.
AFQMC has been successfully applied in recent years to a number of challenging problems in both quantum chemistry\citep{motta_review,zhang_lecture,purwanto_excited_c2,borda2018non,shee_trans} and solid state physics\citep{purwanto_ca_centers,ma_ph_excite,zhang_nio}. However, the broad applicability of the method is not as well understood as more traditional quantum chemistry approaches which have seen decades worth of sustained development and benchmarking.
The primary limiting factor of AFQMC is the choice of trial wavefunction.
Single-determinant trial wavefunctions from Hartree--Fock (HF) or density functional theory calculations have shown remarkable accuracy for a broad range of applications including the two-dimensional Hubbard model\citep{simons_hubbard_2d}, dipole-bound anions\citep{hongxia_anions} and solid state applications\citep{afqmc_pwpsp,afqmc_mppsp,ma_dwnf_prl,purwanto_dwnf_fold_jctc}, with total energies often approaching the accuracy of coupled cluster singles and doubles with perturbative triples CCSD(T)\citep{motta_thc}.
However, for more strongly correlated systems such as transition metal containing complexes single determinant trial wavefunctions are not sufficiently accurate and multi determinant trial wavefunctions become necessary\citep{purwanto_cr2,shi_symm_afqmc,shi_symm_proj,chang2017multi,borda2018non,shee_trans}.

Often short determinantal expansions from complete active space self-consistent field (CASSCF) wavefunctions\citep{purwanto_excited_c2,shee_trans} or non-orthogonal multi-Slater deteraminant trial wavefunction\citep{borda2018non} can help to restore the accuracy of the method.
However, since multi-determinant wavefunctions scale exponentially with system size, this approach to improving the trial wavefunction is ultimately limited, particularly for large scale applications.
The search for more economical and accurate trial wavefunctions for AFQMC (and also FN-DMC) is an active area of research and no single approach can achieve both polynomial scaling and broadly consistent accuracy at the same time.

Our goal is to assess the quality of HF trial wavefunctions for the Uniform Electron Gas (UEG) model\citep{vignale_qtel}.
HF wavefunctions are the simplest possible reference states to perform subsequent correlation calculations in quantum chemistry.
Often, artificial symmetry breaking caused by HF wavefunctions causes confusion in understanding electron correlation.\cite{Jimenez-Hoyos2014,lee2019distinguishing}
\revrep{For instance, with artificial symmetry breaking, subsequent correlation calculations become often ineffective\cite{Jackels1976,Davidson1983,Andrews1991,Ayala1998,McLean1985,Sherrill1999,Crawford2000,Paldus2007}.}{
It is artificial because the symmetry breaking occurs due to the lack of treatment for weak correlation, not strong correlation.
Artificial symmetry breaking affects the performance of subsequent correlation calculations greatly (especially for weakly correlated systems)\cite{Jackels1976,Davidson1983,Andrews1991,Ayala1998,McLean1985,Sherrill1999,Crawford2000,Paldus2007}.
In such cases, it is preferred to use spin-restricted HF (RHF) orbitals as opposed to variationally preferred broken symmetry HF orbitals.}
Therefore, one has to be cautious when choosing a proper HF state for correlation calculations.
Nonetheless, HF is not only the simplest but also scalable to hundreds of electrons.
With HF trials (which we name as HF+ph-AFQMC), AFQMC scales strictly as $\mathcal O(N^3)-\mathcal O(N^4)$ for a fixed statistical accuracy.
Therefore, it is crucial to assess the accuracy of HF+ph-AFQMC and understand the scope of it in simulating large-scale chemical and solid-state systems.
After all, the gold-standard quantum chemistry method, CCSD(T)\cite{raghavachari1989fifth} is performed on top of 
HF states.

In the context of this paper, the UEG model provides us with a simplified version of the full ab-initio Hamiltonian for a solid, essentially omitting the electron-ion interaction term and all of the material complications it entails.
One can tune the magnitude of dynamic and static correlation at the Hamiltonian level using a single parameter (for a fixed number of electrons) through the dimensionless Wigner-Seitz radius $r_s$.
\insertnew{
For low $r_s$ values, the electron density is high, the distance between electrons is short, and electrons are all paired.
In this regime, diagrammatic resummation techniques such as random phase approximation (RPA) and low-order perturbation theory
based on RHF references are highly accurate \cite{ma1968rpa}. We call this regime ``weakly correlated''.
On the other hand, for high $r_s$ values, electrons are unpaired and spatially well-separated and closed-shell RHF references
provide qualitatively wrong picture (i.e., it does not describe the open-shell nature of the system).
In this case, RPA or other low-order perturbation theory based on RHF references fails.
Furthermore, the use of broken symmetry HF references does not provide accurate description either due to spin contamination.
We call this regime ``strongly correlated''.} This tunability allows for a unambiguous comparison between the strengths and weaknesses of various methods.
Moreover, there exist a number of benchmark results both for intermediate system sizes within the reach of traditional quantum chemistry approaches, as well as results for much larger system sizes, and also results extrapolated to the thermodynamic limit. 

Recently, the formally exact full configuration interaction quantum Monte Carlo method\citep{booth_fciqmc,cleland_initiator} (FCIQMC) provided benchmark results for a range of densities for a 14- and 54-electron system\citep{shephed_ueg,shepherd_ueg_jcp}. 
These FCIQMC studies also motivated recent coupled-cluster Monte Carlo\citep{thom_ccmc,spencer_ccmc,franklin_ccmc} studies on the 14-electron UEG model by Neufeld and Thom where
they provided CCSD, CCSD and triples (CCSDT), CCSDT and quadruples (CCSDTQ) with RHF references results for a wide range of $r_s$ and basis sets\citep{neufeld_ccmc}.
As the scope of truncated spin-restricted CC (RCC) approaches is relatively well understood, comparing HF+ph-AFQMC against these results will lead us to a better understanding of the scope of HF+ph-AFQMC.
Although the ph-AFQMC has been applied to 3D UEG before to construct the KZK functional\citep{KZK2008,KZK2011} and to small 2D UEG models\citep{motta_itcf_1,motta_itcf_2}, we believe this is the first published extensive benchmarking study of the 3D UEG using ph-AFQMC. 

This paper is organized as follows: (1) we briefly review the formalism of ph-AFQMC and the UEG model, 
(2) we analyze the basis set convergence of AFQMC in the 14-electron UEG model
and compare its result against FCIQMC and CCMC
, 
and (3) we study 
larger systems (54-electron and 114-electron)
and discuss the AFQMC perspectives on simulating the ground state of solids.

\section{Methods}
In this section we briefly summarize the basics of the AFQMC method and the phaseless approximation which leads to the phaseless AFQMC algorithm (ph-AFQMC).
\insertnew{
We use $n_\text{occ}$ to denote the number of occupied molecular orbitals (MOs) and $n_\text{vir}$ to denote the number of
unoccupied MOs.
}

\subsection{AFQMC}
\subsubsection{Free-Projection AFQMC}
The zero-temperature AFQMC algorithm is a stochastic realization of power methods that target the lowest root of the Hamiltonian $\hat{H}$.
The algorithm is based on the following identity:
\insertnew{
\begin{equation}
|\Psi_0\rangle 
\propto
\lim_{\tau\rightarrow \infty}    
\exp{\left(-\tau \hat{H}\right)} |\Phi_0\rangle
= 
\lim_{\tau\rightarrow \infty}    
|\Psi(\tau)\rangle,
\label{eq:exact}
\end{equation}
}
where
$|\Psi_0\rangle$ is the exact ground state 
and $|\Phi_0\rangle$ is an initial starting wavefunction satisfying $\langle\Phi_0|\Psi_0\rangle \ne 0$. Although the initial wavefunction $|\Psi_0\rangle$ can differ from the trial wavefunction $|\Psi_T\rangle$, for the purpose of this work we will assume 
\revrep{$|\Psi_0\rangle = |\Phi_T\rangle$ when using a RHF trial wavefunction}{$|\Psi_0\rangle = |\Phi_T\rangle$ unless mentioned otherwise}.
Eq. \eqref{eq:exact} is implemented stochastically by repeatedly applying a propagator,
$\exp(-\Delta\tau \hat{H})$,
to a set of random walkers until the ground state is reached.
Each walker is comprised of a Slater determinant, $|\psi_n(\tau)\rangle$, and a weight $w_n(\tau)$ such that the statistical representation of the wavefunction is given by $|\Psi(\tau)\rangle=\sum_n w_n(\tau)|\psi_n(\tau)\rangle$.

In order to practically realize the projection, we first split the Hamiltonian into one-body and two-body operators (i.e., $\hat{H} = \hat{H}_1+\hat{H}_2$).
For the two-body terms,  we write them in the sum of squared operators,
\begin{equation}
    \hat{H}_2 = -\frac12 \sum_\alpha^{N_\alpha} \hat{v}_\alpha^2.
\end{equation}
Then,
we apply the Hubbard-Stratonovich\citep{hubbard_strat} transformation to rewrite the imaginary-time propagator in terms of only one-body operators. With the symmetric Trotter decomposition, the propagator reads
\begin{equation}
\exp(-\Delta\tau \hat{H}) \: = 
\int d\mathbf{x}
p(\mathbf{x})
\hat{B}(\Delta \tau, \mathbf x),
\label{eq:HS}
\end{equation}
where $p(\mathbf{x})$ is the standard normal distribution, $\mathbf x$ is a vector of $N_\alpha$ auxiliary fields and $\hat{B}$ is defined as
\begin{equation}
\hat{B}(\Delta \tau, \mathbf x) = e^{-\frac{\Delta\tau}{2} \hat{H}_1}
e^{-\sqrt{\Delta\tau} \mathbf{x}\cdot\hat{\mathbf{v}}}
e^{-\frac{\Delta\tau}{2} \hat{H}_1}.
\label{eq:B}
\end{equation}
At each time step, each walker draws Gaussian random numbers to sample one instance of $\mathbf x$ and provides a sample to the HS transformation in Eq. \eqref{eq:HS}. 
The application of a one-body operator such as \cref{eq:B} to a Slater determinant yields yet another single  Slater determinant\citep{thouless_theorem,thouless_theorem_2}. 

For a generic ab-initio Hamiltonian the propagators appearing in \cref{eq:B} will in general be complex and the weights of the walkers will acquire a phase that will be distributed uniformly in the complex plane in the long imaginary time limit\citep{zhang_lecture}.
This `phase problem' is analogous to the notorious fermion sign problem encountered in DMC and has no known solution in general. 
The phase problem can be somewhat mitigated through mean-field subtraction\citep{alsaidi_gaussian} (i.e., redefining $\hat{v}_\alpha' = \hat{v}_\alpha - \langle\hat{v}_\alpha\rangle_0$) in Eq. \eqref{eq:B}, but the statistics will be eventually swamped by the phase problem. Note that mean field subtraction is essentially identical to normal-ordering
$\hat{v}_\alpha$ which ensures $\hat{v}_\alpha' |\Phi_0\rangle = 0$ for all $\alpha$.

\subsubsection{Phaseless AFQMC}
It is possible to eliminate this phase problem entirely at the sake of introducing biases into the results using the so-called phaseless approximation\citep{zhang_phaseless}. This is achieved by first performing an importance sampling transformation to the propagator such that walkers now undergo the modified propagation:
\begin{align}\nonumber
&w_n(\tau+\Delta\tau) |\psi_n(\tau+\Delta\tau)\rangle =\\
&\left[
I(\mathbf{x}_n,\mathbf{\bar{x}}_n,\tau,\Delta\tau)
\hat{B}(\Delta \tau, \mathbf {x}_n-\mathbf{\bar{x}}_n)
\right] w_n(\tau)|\psi_n(\tau)\rangle,
\end{align}
where the importance function (in hybrid form) is defined as 
\begin{equation}
    I(\mathbf{x}_n,\mathbf{\bar{x}}_n,\tau,\Delta\tau) = S_n(\tau, \Delta\tau)
    e^{\mathbf{x}_n\cdot\mathbf{\bar{x}}_n-\mathbf{\bar{x}}_n\cdot\mathbf{\bar{x}}_n/2},
 \label{eq:import}
\end{equation}
$S_n$ is the overlap ratio of the $n$-th walker
\begin{equation}
S_n(\tau, \Delta\tau) = \frac{\langle
    \Psi_T |
    \hat{B}(\Delta\tau, \mathbf{x}_n-\mathbf{\bar{x}}_n) | \psi_n(\tau)
    \rangle}{
    \langle
    \Psi_T | \psi_n(\tau)
    \rangle},
 \label{eq:ovl}
\end{equation}
%$|\Psi_T\rangle$ is the trial wavefunction 
and $\mathbf{\bar{x}}_n$ is an ``optimal'' force bias which is a shift to the Gaussian distribution, given as
\begin{equation}
    \mathbf{\bar{x}}_n(\Delta \tau, \tau) = -\sqrt{\Delta\tau}
\frac{\langle
    \Psi_T | \hat{\mathbf{v}}' | \psi_n(\tau)
    \rangle}{
    \langle
    \Psi_T | \psi_n(\tau)
    \rangle
    }.
\end{equation}
The phaseless approximation (ph) is then defined as a modification to this importance function
\begin{equation}
I_{ph}(\mathbf{x}_n,\mathbf{\bar{x}}_n,\tau,\Delta\tau) = |I (\mathbf{x}_n,\mathbf{\bar{x}}_n,\tau,\Delta\tau)|\times
\text{max}(0, \cos(\theta_n(\tau)))
\label{eq:ph}
\end{equation}
where the phase $\theta_n(\tau)$ is given by
\begin{equation}
\theta_n(\tau) = \text{arg}\left(
%\frac{\langle
%    \Psi_T | \hat{B}(\Delta\tau, \mathbf{x}-\mathbf{\bar{x}}) | \Psi(\tau)
%    \rangle}{
%    \langle
%    \Psi_T | \Psi(\tau)
%    \rangle
%    }
S_n(\tau, \Delta\tau)
\right).
\label{eq:theta}
\end{equation}
The walker weights and Slater determinants are then updated as 
\begin{align}
w_n(\tau+\Delta\tau) &= I_{ph}(\mathbf{x}_n,\mathbf{\bar{x}}_n,\tau,\Delta\tau) \times w_n(\tau)\\
|\psi_n(\tau+\Delta\tau)\rangle &= \hat{B}(\Delta\tau, \mathbf{x}_n-\mathbf{\bar{x}}_n) |\psi_n(\tau)\rangle. 
\end{align}
Evidently, the phaseless approximation ensures that the walker weights remain real and non-negative throughout the simulation and therefore removes the phase problem completely.

The mixed estimate for the local energy estimator can be computed with the generalized Green's function (or one-particle reduced density matrix) $\mathbf P$,
\begin{equation}
P_{pq} = 
\frac{
\langle \Psi_T
|
\hat{a}_p^\dagger\hat{a}_q
|\psi_n(\tau)\rangle
}
{
\langle \Psi_T
|\psi_n(\tau)\rangle
}
= 
\left(\mathbf {C}_{\psi_n}
\left(
\mathbf {C}_{\Psi_T}^\dagger
\mathbf {C}_{\psi_n}
\right)^{-1}
\mathbf {C}_{\Psi_T}^\dagger\right)_{qp}
\label{eq:rdm}
\end{equation}
where $\mathbf {C}_{\psi_n}$ is 
the occupied \revrep{molecular orbital}{MO} coefficient of $|\psi_n(\tau)\rangle$
and $\mathbf {C}_{\Psi_T}$ the occupied \revrep{molecular orbital}{MO} coefficient of $|\Psi_T\rangle$.
Once the simulation has equilibrated, we will have a statistical representation of the ground state wavefunction given by
\begin{equation}
    |\Psi(\tau)\rangle = \sum_n w_n(\tau) \frac{|\psi_n(\tau)\rangle}{\langle\Psi_T|\psi_n(\tau)\rangle},
\end{equation}
from which we can compute the mixed estimator for the energy as
\begin{equation}
E(\tau) = 
\frac{\langle \Psi_T
|
\hat{H}
| \Psi(\tau)\rangle}{
\langle \Psi_T
| \Psi(\tau)\rangle
}
=
\frac{
\sum_n w_n (\tau) \epsilon_n(\tau)
}
{
\sum_n w_n (\tau) 
},
\label{eq:mixedE}
\end{equation}
where  $\epsilon_n(\tau)$ is the local energy of a walker,
\insertnew{
\begin{equation}
\epsilon_n(\tau) = 
\frac{\langle \Psi_T | \hat{H} | \psi_n(\tau)\rangle}{\langle \Psi_T |  \psi_n(\tau)\rangle}.
\end{equation}
}
We will see how the local energy evaluation is done specifically for the UEG model later.

\subsubsection{Size-consistency of ph-AFQMC}\label{sec:size}
Size-consistency is a property of a wavefunction
for isolated systems $A$ and $B$
 that asserts the product separability of a supersystem
wavefunction ($|\Psi_{AB}\rangle = |\Psi_A\rangle|\Psi_B\rangle$) and also the additive separability of energy ($E_{AB} = E_A+E_B$).
Configuration interaction (CI) based quantum chemistry methods are in general not size-consistent.\cite{bartlett1977determination}
In particular, the only size-consistent CI methods are CI with singles (CIS) and FCI.
On the other hand, single-reference CC methods are size-consistent as long as the form of wavefunction is parametrized by an exponential of the 
cluster operator. 
\revrep{This is why CC methods are promising in terms of simulating solids because
size-consistency is crucial to reliably obtain energy of large systems including solids.}{
To reliably obtain the thermodynamic limit of large systems, 
size-consistency is necessary. At the thermodynamic limit, size-inconsistent methods
approach just mean-field total energy and estimate no correlation energy.
Therefore, CC methods, due to size-consistency, have stood out as a unique
tool for simulating bulk systems \cite{Gruber2018}.}

We will show that ph-AFQMC is also size-consistent as long as the trial wavefunction is product separable. 
For isolated systems $A$ and $B$, the supersystem Hamiltonian separates into $\hat{H}_A$ and $\hat{H}_B$.
Furthermore, these two operators commute since these systems are isolated.
Therefore, the propagator is also product separable,
\begin{equation}
\exp(-\Delta\tau \hat{H}_{AB})
=
\exp(-\Delta\tau \hat{H}_{A})
\exp(-\Delta\tau \hat{H}_{B})
%\hat{B}_{AB}
%=
%\hat{B}_{A}
%\hat{B}_{B}
\end{equation}
The HS transformation can be performed on $\exp(-\Delta\tau \hat{H}_{A})$ and $\exp(-\Delta\tau \hat{H}_{B})$
separately so that we have
$
\hat{B}_{AB}
=
\hat{B}_{A}
\hat{B}_{B}.
$
This proves the size-consistency of free-projection AFQMC.

It can be also shown that the phaseless constraint is product separable.
The overlap function in Eq. \eqref{eq:ovl} can be written as
\begin{equation}
S_n^{AB} = 
\frac{\langle
    \Psi_T^A | 
    \hat{B}_A | \psi_n^A
    \rangle}{
    \langle
    \Psi_T^A | \psi_n^A
    \rangle}
\frac{\langle
    \Psi_T^B | 
    \hat{B}_B | \psi_n^B
    \rangle}{
    \langle
    \Psi_T^B | \psi_n^B
    \rangle}
    = S_n^AS_n^B
\end{equation}
where
the only assumptions we are making are
(1) the product separability of the trial wavefunction: $|\Psi_T^{AB}\rangle = |\Psi_T^{A}\rangle|\Psi_T^{B}\rangle$
and
(2) the product separability of the slater determinant of $n$-th walker:
$|\psi_n^{AB}\rangle = |\psi_n^{A}\rangle|\psi_n^{B}\rangle$.
The assumption (2) can be satisfied as long as we start from a product separable wavefunction since
the propagator is product separable.
With this overlap function, one can show that the importance function also obeys the product separability
and therefore we conclude that ph-AFQMC is {\it size-consistent}.
\subsection{Uniform Electron Gas}
The Hamiltonian for the uniform electron gas (UEG) is given simply as the sum of the kinetic energy and electron-electron interaction operator (up to a constant):
\begin{equation}
\hat{H} = \hat{T} + \hat{V}_\text{ee} + E_M.
\end{equation}
We will work with a basis of planewave spin orbitals $\langle \mathbf{r}\sigma|\mathbf{G}_i\sigma_i\rangle = \frac{1}{L^{3/2}} e^{i\mathbf{G}_i\cdot\mathbf{r}}\delta_{\sigma,\sigma_i}$, where $L$ is the length of the simulation cell, $\mathbf{G}_i=\frac{2\pi}{L}\mathbf{n}_i$ for $\mathbf{n}_i$ a vector of integers and $\sigma_i$ is a spin index (either $\alpha$ or $\beta$).
We impose a kinetic energy cutoff $E_{\text{cut}}$ and work with a finite basis of $2M$ spin orbitals.
In this basis the kinetic energy is written as
\begin{equation}
\hat{T} = \sum_{\mathbf G} \frac{|\mathbf G|^2}{2} a_{\mathbf G}^\dagger a_{\mathbf G},
\end{equation}
and the electron-electron interaction operator is given by
\begin{equation}
\hat{V}_\text{ee} = \frac{1}{2\Omega	}
%\sum_{\sigma_1, \sigma_2}
\sum_{\mathbf{Q}\ne\mathbf 0,\mathbf{G}_1,\mathbf{G}_2}
\frac{4\pi}
{|\mathbf{Q}|^2}
a_{\mathbf{G}_1+\mathbf{Q}}^\dagger
a_{\mathbf{G}_2-\mathbf{Q}}^\dagger
a_{\mathbf{G}_2}
a_{\mathbf{G}_1},
%a_{\mathbf{G}_1+\mathbf{Q},\sigma_1}^\dagger
%a_{\mathbf{G}_2-\mathbf{Q},\sigma_2}^\dagger
%a_{\mathbf{G}_2,\sigma_2}
%a_{\mathbf{G}_1,\sigma_1}
\end{equation}
where $\Omega = L^3$ is the simulation cell volume, $\mathbf{Q}$ is a momentum transfer vector that lives in an enlarged basis of size $4E_\text{cut}$ and we have dropped the subscript index on $\mathbf{G}$ for simplicity.
Lastly, the Madelung energy $E_M$ is included to account for the self-interaction of the Ewald sum under periodic boundary conditions\citep{fraser_fzc}.
For simplicity we use the formula proposed by Schoof and co-workers\citep{schoof_prl}
\begin{equation}
E_M \approx -2.837297 \times \left(\frac{3}{4\pi}\right)^{1/3}N^{2/3}r_s^{-1},
\end{equation}
where $N$ is the number of electrons in the simulation cell cell and $r_s = \left(\frac{3L^3}{4\pi N}\right)^{1/3}$ is the dimensionless Wigner-Seitz radius.

The local energy $\epsilon_n(\tau)$ for the UEG then reads
\begin{align}\nonumber
\epsilon_n(\tau)
&=
E_M + 
\sum_{\mathbf G} \frac{|\mathbf G|^2}{2} P_{\mathbf G \mathbf G}\\
&+
\frac{1}{2\Omega}\sum_{\mathbf{Q}\ne\mathbf 0}
\frac{4\pi}
{|\mathbf{Q}|^2} \left(
\Gamma_{\mathbf Q}
-
\Lambda_{\mathbf{Q}}\right),
%\\
%&
%-
%\frac{1}{2\Omega}
%\sum_{\sigma_1, \sigma_2}
%\sum_{\mathbf{G}\ne\mathbf 0,\mathbf{G}_1,\mathbf{G}_2}
%\frac{4\pi}
%{|\mathbf{G}|^2}
%\Lambda_{\mathbf{G}}
%P_{\mathbf{G}_1+\mathbf{G},\mathbf{G}_2}
%P_{\mathbf{G}_2-\mathbf{G},\mathbf{G}_1}
%
\end{align}
where the Coulomb two-body density matrix $\Gamma_{\mathbf Q}$ is
\begin{equation}
\Gamma_{\mathbf Q} = 
\left(\sum_{\mathbf G_1}P_{\mathbf{G}_1+\mathbf{Q},\mathbf{G}_1}\right)
\left(\sum_{\mathbf G_2}P_{\mathbf{G}_2-\mathbf{Q},\mathbf{G}_2}\right)
\end{equation}
and the exchange two-body density matrix $\Lambda_{\mathbf Q}$ is
\begin{equation}
\Lambda_{\mathbf{Q}} = 
\sum_{\mathbf{G}_1\mathbf{G}_2}
P_{\mathbf{G}_1+\mathbf{Q},\mathbf{G}_2}
P_{\mathbf{G}_2-\mathbf{Q},\mathbf{G}_1}
\end{equation}
The formation of $\Gamma_{\mathbf Q}$ costs $\mathcal O(M^2)$ whereas the formation of $\Lambda_{\mathbf Q}$ takes $\mathcal O(M^3)$ amount of work.
Therefore, the evaluation of the exchange contribution is the bottleneck in the local energy evaluation.
As noted in Ref.~\citenum{afqmc_pwpsp} the evaluation of the energy (and propagation) can be accelerated using fast Fourier transforms, however we did not use this optimization here.

The two-body Hamiltonian $\hat{V}_\text{ee}$ needs to be rewritten as a sum of squares to employ
the AFQMC algorithm. It was shown in Ref.~\citenum{afqmc_pwpsp} that
\begin{equation}
\hat{V}_\text{ee} = \frac14 \sum_{\mathbf Q \ne \mathbf 0}
\left[
\hat{A}^2(\mathbf Q)
+
\hat{B}^2(\mathbf Q)
\right],
\end{equation}
where
\begin{equation}
\hat{A}(\mathbf Q) = \sqrt{\frac{2\pi}{\Omega |\mathbf Q|^2}}
\left(
\hat{\rho}(\mathbf Q)
+ \hat{\rho}^\dagger(\mathbf Q)
\right),
\end{equation}
and
\begin{equation}
\hat{B}(\mathbf Q) = i\sqrt{\frac{2\pi}{\Omega |\mathbf Q|^2}},
\left(
\hat{\rho}(\mathbf Q)
- \hat{\rho}^\dagger(\mathbf Q)
\right),
\end{equation}
with the momentum transfer operator $\hat{\rho}$ defined as
\begin{equation}
\hat{\rho}(\mathbf Q) = 
%\sum_\sigma 
\sum_{\mathbf G} a^\dagger_{
\mathbf G + \mathbf Q%, \sigma
}
a_{
\mathbf G%, \sigma
} 
\Theta\left(E_\text{cut} - \frac{|\mathbf G + \mathbf Q|^2}{2}\right),
\end{equation}
where
$\Theta$ is the Heaviside step function.
The Hubbard-Stratonovich operators $\hat{\mathbf v}$ are now $\hat{A}(\mathbf Q)$ and $\hat{B}(\mathbf Q)$, and the rest of the AFQMC algorithm follows straightforwardly.
\subsection{Hartree-Fock Trial Wavefunctions}\label{subsec:hftrial}
In ph-AFQMC, the main source of error is the bias introduced by the phaseless constraint. The magnitude of this bias is heavily dependent on the quality of trial wavefunctions.
Although there are advanced options available for these such as multideterminantal trials and self-consistently determined single-determinantal trials\citep{qin_self_cons},
we will employ a simple single determinant RHF trial wavefunction in most cases.
In the UEG model, this is an $M\times N$ matrix (where $N$ is the number of electrons and $M$ is the number of planewaves) with 1's on the diagonal entries.

Typically, for strongly correlated systems it is useful to exploit essential symmetry breaking with HF wavefunctions.
It is essential (as opposed to artificial) in the sense that the property of a single determinant wavefunction is qualitatively wrong without it.
An attempt to exploit \revrep{artificial}{essential} symmetry breaking typically leads to either spin-unrestricted HF (UHF) or spin-generalized HF (GHF) \insertnew{trial wavefunctions} which have a lower energy than RHF.
\insertnew{Indeed, such essential symmetry breaking was shown to be powerful when applying ph-AFQMC to bond dissociation of F$_2$ \cite{purwanto2008eliminating}.}

\insertnew{An example of artificial symmetry breaking is buckminsterfullerene (C$_{60}$), a stable, electron paramagnetic resonance silent (EPR-silent) molecule \cite{Paul2002}.
There is a complex, GHF (cGHF) solution \cite{Jimenez-Hoyos2014} for C$_{60}$ which was characterized to be an artifact due to the lack of treatment for weak correlation at the HF level \cite{lee2019distinguishing}. 
In other words, orbital optimization in the presence of weak correlation such as the second-order M{\o}ller-Plesset theory
would restore artificial symmetry breaking.
Since both experiments \cite{Beckhaus1992,Tomita2003} and computations\cite{lee2019distinguishing} suggest that C$_{60}$ is a stable closed-shell molecule,
the RHF state is more qualitatively correct than other broken-symmetry HF states.
A detailed study of artificial versus essential symmetry breaking in ph-AFQMC will be published in our forthcoming paper.}

The instability of RHF solutions is expected at all $r_s$ values of the UEG model at the thermodynamic limit as proven by Overhauser.\cite{overhauser1960giant,overhauser1962spin,overhauser1968exchange}
As mentioned in Ref.\citenum{zhang_ueg_hf}, however, the R to U spin-symmetry breaking may not occur in the UEG model with a finite number of electrons.
Instead, there is a critical Wigner-Seitz radius ($r_s^c$) below which no UHF solution exists.
This is not surprising in the context of quantum chemistry since this is the same concept as ``Coulson-Fischer points''\cite{Coulson1949} in molecules. \insertnew{Namely, when dissociating molecules there exists a critical bond length where the R to U instability occurs. At bond distances closer than this, there is no genuine UHF solution.}

\insertnew{
In order to perform HF calculations, one must compute the effective one-body operator called the ``Fock'' operator defined as
\begin{equation}
\mathbf F = \frac{\partial E}{\partial \mathbf D}
\end{equation}
where $\mathbf D = \mathbf C_\text{occ} \mathbf C_\text{occ}^\dagger$ with $\mathbf C_\text{occ}$ being the occupied MO coefficient matrix.
After some straightforward algebra,
we find
\begin{equation}
\mathbf F = \mathbf T + \mathbf J - \mathbf K
\end{equation}
where
the kinetic energy matrix, $\mathbf T$, reads
\begin{equation}
T_{\mathbf G, \mathbf G'} = \frac 12 |\mathbf G|^2 \delta_{\mathbf G, \mathbf G'},
\end{equation}
the Coulomb matrix, $\mathbf J$, is
\begin{align}
J_{\mathbf G, \mathbf G'}
=
\frac{1}{2\Omega}
\sum_{\mathbf Q\ne0}
\frac{4\pi}{|\mathbf Q|^2}
\frac{\partial \Gamma_{\mathbf Q}}{\partial D_{\mathbf G, \mathbf G'}}
\end{align}
with
\begin{align}\nonumber
\frac{\partial \Gamma_\mathbf Q}{\partial D_{\mathbf G, \mathbf G'}} 
& = \left(\sum_{\mathbf G_1}\delta_{\mathbf G, \mathbf{G}_1+\mathbf{Q}}\delta_{\mathbf G', \mathbf{G}_1}\right)
\left(\sum_{\mathbf G_2}D_{\mathbf{G}_2-\mathbf{Q},\mathbf{G}_2}\right) \\
&+
\left(\sum_{\mathbf G_1}D_{\mathbf{G}_1+\mathbf{Q},\mathbf{G}_1}\right)
\left(\sum_{\mathbf G_2}\delta_{\mathbf G, \mathbf{G}_2-\mathbf{Q}}\delta_{\mathbf G'\mathbf{G}_2}\right)
\end{align}
and the exchange matrix is given by
\begin{align}
K_{\mathbf G, \mathbf G'}
=
\frac{1}{2\Omega}
\sum_{\mathbf Q\ne0}
\frac{4\pi}{|\mathbf Q|^2}
\frac{\partial \Lambda_{\mathbf Q}}{\partial D_{\mathbf G, \mathbf G'}}
\end{align}
with
\begin{align}\nonumber
\frac{\partial \Lambda_\mathbf Q}{\partial D_{\mathbf G, \mathbf G'}} 
& = \sum_{\mathbf{G}_1\mathbf{G}_2}\bigg(
\delta_{\mathbf G, \mathbf{G}_1+\mathbf{Q}}\delta_{\mathbf G',\mathbf{G}_2}
D_{\mathbf{G}_2-\mathbf{Q},\mathbf{G}_1}\\
&+
D_{\mathbf{G}_1+\mathbf{Q},\mathbf{G}_2}
\delta_{\mathbf G, \mathbf{G}_2-\mathbf{Q}}\delta_{\mathbf G', \mathbf{G}_1}
\bigg)
\end{align}
A similar derivation can be found in Ref. \citenum{zhang_ueg_hf}.
}

\insertnew{
With the above Fock matrix, one can perform an HF calculation by
optimizing the HF energy expression with respect to the orbital rotation parameter
$\Theta_\text{vo}$ (a matrix of $n_\text{vir}$-by-$n_\text{occ}$) which relates
two different MO coefficients via a unitary transformation,
\begin{equation}
\mathbf C' = \mathbf C \exp\left(\mathbf \Delta\right)
\end{equation}
where the antihermition matrix $\Delta$, which is parametrized by $\Theta_{vo}$,
\begin{equation}
\mathbf \Delta = 
\begin{bmatrix} 
\mathbf 0_\text{oo} & -\mathbf \Theta_\text{vo}^\dagger \\
\mathbf \Theta_\text{vo} & \mathbf 0_\text{vv}
\end{bmatrix}
\end{equation}
The subscript of each matrix block denotes the dimension of the corresponding block, $o = n_\text{nocc}$ and 
$v = n_\text{nvir}$.
An HF solution is defined as a stationary point that satisfies
\begin{equation}
\frac{\partial E_\text{HF}}{\partial \mathbf\Theta_\text{vo}} = 0
\end{equation}
where 
\begin{equation}
E_\text{HF} = 
\langle\Phi_\text{HF}|
\hat{T} + \hat{V}_\text{ee}
|
\Phi_\text{HF}\rangle.
\end{equation}
The local stability of a given stationary point can then be tested by diagonalizing the orbital Hessian, $\mathbf M$,
\begin{equation}
M_{ai,bj} = \frac{\partial^2 E}{\partial\Theta_{ai}\partial\Theta_{bj}}
\end{equation}
}
%Hartree-Fock calculations solve the following non-linear equation:
%\begin{equation}
%\mathbf F \mathbf C = \mathbf C \mathbf \epsilon
%\end{equation}
%where $\mathbf C$ is the molecular orbital (MO) coefficient matrix, $\mathbf \epsilon$ is the MO energies, and $\mathbf F$ is the Fock matrix defined as
%\begin{equation}
%\mathbf F[\mathbf P] = \frac{\partial E}{\partial \mathbf P}
%\end{equation}
%with $\mathbf P$ depends solely on the occupied orbitals (i.e., $\mathbf P = \mathbf C_\text{occ} \mathbf C_\text{occ}^\dagger$).
%Due to the Coulomb and exchange interactions, $\mathbf F $ itself depends on $\mathbf P$ there fore the above eigenvalue equation is non-linear.

\insertnew{
We will see whether there is essential symmetry breaking 
in the low $r_s$ regime in the UEG model and
try to utilize this essential symmetry breaking when appropriate.
}
%Instead of utilizing this essential symmetry breaking of the UEG model, we will
%use RHF trial wavefunctions for most AFQMC calculations and assess its performance.
%It will be interesting to see how much improvement one can achieve with UHF or GHF trial wavefunctions in the future.
%In other words, as long as there is no obvious orbital degeneracy, there is no symmetry breaking at the HF level.
%We utilized such essential symmetry breaking for some cases in conjunction with non-orthogonal configuration interaction (NOCI) expansions and discuss the utility of these trial wavefunctions.
%As shown in Ref. [], the R to U spin-symmetry breaking does not occur in the UEG model when the number of electrons is one of the ``magic numbers''.
%In other words, as long as there is no obvious orbital degeneracy, there is no symmetry breaking at the HF level.

%\subsection{Algorithm/Scaling/Timestep}
\section{Computational Details}
Unless otherwise noted, the AFQMC calculations in this work were performed by a development version of \texttt{QMCPACK}\citep{qmcpack}.
\texttt{PAUXY}\citep{pauxy} was also used in the initial testing stages.
Unless noted otherwise, AFQMC results below are obtained using \texttt{QMCPACK}. \texttt{HANDE}\citep{HANDEdev,hande_jctc} was used to crosscheck our numbers for small systems that are not presented in this work.
We used 0.005 a.u. for the time step $\Delta \tau$ throughout the paper. This was found to be enough for systems we considered here.
A total of 2880 walkers were used and the population bias from this was found to be negligible in the results reported here. 
The comb\citep{booth_comb} and pair branching\citep{wagner_qwalk} population control algorithms are used in \texttt{PAUXY} and \texttt{QMCPACK} respectively. \insertnew{The demonstration of the convergence of these parameters is available in the Supplementary Materials.}

\insertnew{All broken symmetry HF calculations were performed with a development version of Q-Chem \cite{Shao2015} and the details for the implementation can be found in refs. \citenum{lee2018regularized,lee2019distinguishing,lee2019two}. The optimizer used for those HF calculations is geometry direct minimization (GDM) developed by Van Voorhis and Head-Gordon \cite{Voorhis2002}.  The internal stability analysis was performed for all HF solutions to ensure the local stability of each solution \cite{Seeger1977} where we used a finite-difference orbital Hessian using analytic orbital gradient \cite{Sharada2015}.} All calculations were performed with periodic boundary conditions; no twist averaging was performed.

\section{Results}
%\subsection{Zero Temperature}
The UEG model has been explored by multiple methods at $T=0$ and an extensive amount of benchmark data are already available.
We compare our ph-AFQMC results against other methods and discuss whether the use of RHF trial wavefunction is reliable for $r_s \le 5.0$.
It is expected that the quality of an RHF wavefunction degrades as $r_s$ increases (approaching the atomic limit) since electrons tend to localize. 
$r_s=5.0$ is a commonly investigated intermediate Wigner-Seitz radius so it will be interesting to see how ph-AFQMC performs without employing more sophisticated trial wavefunctions.

For simplicity, we will refer ph-AFQMC with an RHF trial wavefunction (RHF+ph-AFQMC) to as ph-AFQMC \insertnew{unless mentioned otherwise}. 

\insertnew{
\subsection{Broken-Symmetry HF States}
\label{subsec:uhf}
We summarize some interesting aspects in the HF solutions of the UEG model due to its simple form of Hamiltonian:
\begin{enumerate}
\item The MO coefficient matrix of an RHF solution is an identity matrix.
This makes obtaining an UHF solution from solving an eigenvalue equation for $\mathbf F$ difficult in the following sense.
$\mathbf C$ from diagonalizing $\mathbf F$ is always unitary and the subsequent density matrix $\mathbf D$ is therefore identity.
Since $\mathbf D$ is identical to the density matrix of RHF, one obtains an RHF solution immediately after one single diagonalization of $\mathbf F$. A direct energy minimization \cite{Voorhis2002} or the use of the HF projector \cite{zhang_ueg_hf} is necessary to obtain broken-symmetry HF states.
\item The RHF energy does not depend on the basis set size.
However, broken-symmetry solutions depend on the basis set size.
It is important to converge their energies with respect to $M$ when discussing their existence.
\item Both Coulomb and kinetic energies are {\it minimized} in an RHF solution.
In particular, the Coulomb energy is {\it always} zero in RHF.
\item The R to U symmetry breaking is driven by
the lowering of exchange energy which 
compensates the increase in
the Coulomb and kinetic energies.
\end{enumerate}
We are interested in the paramagnetic phase of the UEG.
As $r_s$ increases, the ferromagnetic (i.e., spin-polarized) phase becomes the ground state\cite{Zong2002}.
The GHF solution can appear in the transition between these two phases at quite high $r_s$ values.
Other than this transition, a genuine GHF solution does not appear
and therefore we study only the UHF solutions for the purpose of this study.
}

\begin{figure*}[ht]
\includegraphics[scale=0.7]{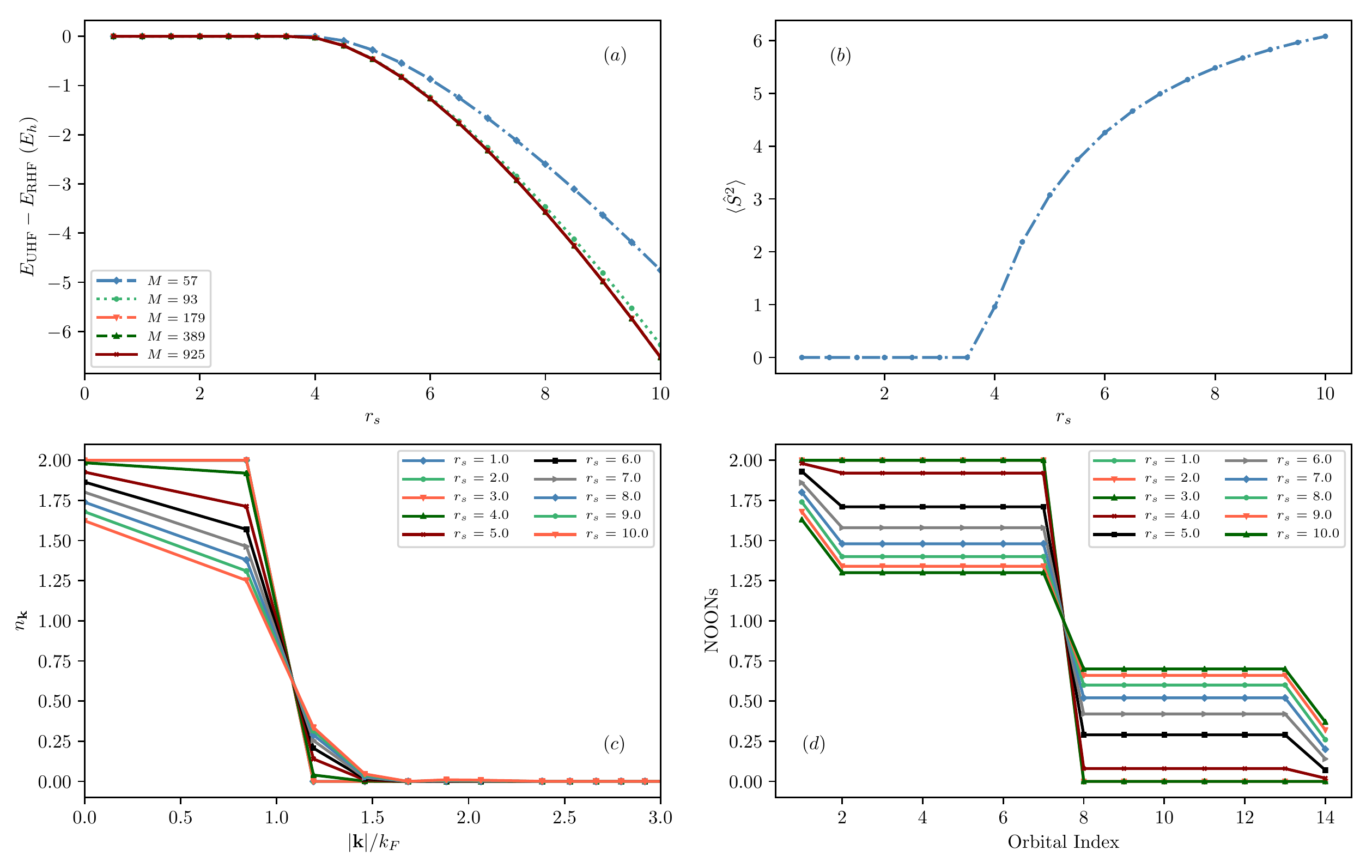}
\caption{
Results of UHF solutions found in the 14-electron UEG model:
(a) the basis set convergence behavior of the energy lowering from RHF to UHF with $M= 57, 93, 179, 389, 925$ over $r_s \in [0.5, 10.0]$,
(b) $\langle \hat{S}^2 \rangle $ as a function of $r_s$ with $M = 925$,
(c) the momentum distribution $n_{\mathbf k}$ for various $r_s$ with $M = 925$,
and
(d) the natural orbital occupation numbers (NOONs) for various $r_s$ with $M = 925$.
In (a), $M=389$ and $M=925$ are more or less on top of each other and visually indistinguishable.
In (c) and (d), curves for $r_s \le 3.0$ are exactly on top of each other as their occupation numbers are either 2 or 0.
%The basis set convergence in the ph-AFQMC correlation energy, $E_c$, of the 14-electron UEG model at $r_s=0.5$.
%The basis sets considered are $M= 57, 93, 179, 389, 925, 1189, 1213, 1419, 2109$.
%The ph-AFQMC values are reproduced in \cref{tab:rs05} for clarity.
%The inset shows the correlation energy for $M\ge 389$.
%Note that the onset of a clear $1/M$ dependence occurs at least past $M=925$.
}
\label{fig:uhf}
\end{figure*}

\insertnew{
We discuss the UHF solutions in the 14-electron UEG model.
In \cref{fig:uhf} (a), 
we show the basis set convergence behavior of UHF.
Unlike RHF, the UHF energy does depend on the basis set size.
For the 14-electron UEG model, it is sufficient to converge the 
UHF energies over $r_s \le 10.0$ with $M = 925$.
Based on \cref{fig:uhf} (a), 
we see that the energy lowering from RHF to UHF
starts to appear for $r_s > 3.5$.
The critical Wigner-Seitz radius for the 14-electron model is $r_s^c \in (3.5,4.0]$.
This is more obvious from looking at the 
$\langle \hat{S}^2 \rangle$ value of the UHF solutions for $M=943$ as a function of $r_s$ as shown in
\cref{fig:uhf} (b).
Non-zero $\langle \hat{S}^2 \rangle$ indicates
the appearance of a UHF solution.
It is clear that the RHF solution becomes unstable above $r_s = 3.5$.
The emergent strong correlation as increasing $r_s$ is
most obvious 
from looking at the momentum distribution (MD) (\cref{fig:uhf} (c)) and natural orbital occupation numbers (NOONs) (\cref{fig:uhf} (d)).
The MD is defined as the diagonal elements of a one-particle reduced density matrix,
\begin{equation}
n_\mathbf{k} = \langle a_{\mathbf k_\alpha}^\dagger a_{\mathbf k_\alpha} + a_{\mathbf k_\beta}^\dagger a_{\mathbf k_\beta} \rangle
\end{equation}
The MD for the UEG model was throughly studied in Ref. \citenum{Holzmann2011}.
NOONs are the eigenvalues of
a one-particle density matrix which is closely related to the MD. 
Both MD and NOONs show the increasing number of open-shell electrons as increasing the $r_s$ values.
The open-shell electrons appear for $r_s > 3.5$ which is consistent with 
\cref{fig:uhf} (a) and (b).
}

\insertnew{We also studied larger UEG models, 54-electron and 114-electron, using UHF.
As expected,
there are many more local minima than the 14-electron model.
This makes locating the global minimum even more challenging. 
These multiple minima lead
to ambiguity in the subsequent ph-AFQMC calculations since there are 
many UHF solutions that can be used as trial wavefunctions.
Thorough studies of the UHF solutions in the UEG model were presented in Ref. \citenum{zhang_ueg_hf}.
For the purpose of this work,
instead of trying to locate the global minimum,
we investigated
the critical Wigner-Seitz radius
to determine 
whether one can employ UHF trial wavefunctions for simulating dense UEG models.
We summarize the range for the critical radius for each UEG model studied in this work in \cref{tab:rsc}.}

\insertnew{
Given \cref{tab:rsc}, it is necessary to employ RHF trial wavefunctions 
for the 14-electron model at $r_s \le 3.5$, the 54-electron model at $r_s \le 4.5$, and the 114-electron model at $r_s \le 2.5$.
We will study the 14-electron model at $r_s = 0.5, 1.0, 2.0, 3.0, 5.0$ and investigate the utility of various trial wavefunctions including RHF and UHF at $r_s = 5.0$.
We will focus on using RHF trial wavefunctions for the 54- and 114-electron models at $r_s = 0.5, 1.0, 2.0$ where
there is no R to U instability.
For the 54-electron model at $r_s = 5.0$, there is a UHF solution with $\langle \hat{S} ^2\rangle = 0.94$
which exhibits marginal symmetry breaking.
\begin{table}
  \centering
  \begin{tabular}{|c|c|}\hline
$N$ & Range \\\hline
14 & $r_s^c \in (3.5, 4.0]$ \\ \hline
54 & $r_s^c \in (4.5, 5.0]$ \\ \hline
114 & $r_s^c \in (2.5,3.0]$ \\\hline
%$M$ & ph-AFQMC & $i$-FCIQMC & \multicolumn{1}{c|}{RCCSD} & \multicolumn{1}{c}{RCCSDT} \\ \Xhline{3\arrayrulewidth}
%57 & -0.5173(1) & -0.5169(1) & N/A & N/A \\ \hline
%93 & -0.5592(2) & -0.5589(1) & N/A & N/A \\ \hline
%179 & -0.5794(2) & -0.5797(3) & -0.57365(1) & -0.57971(3) \\ \hline
%389 & -0.5881(2) & -0.5893(3) & N/A & N/A \\ \hline
%925 & -0.5920(8) & -0.5936(3) & -0.58626(4) & -0.5923(1) \\ \hline
%1189 & -0.5921(2) & -0.5939(4) & N/A & N/A \\ \hline
%1213 & -0.5926(8) & N/A & N/A & N/A \\ \hline
%1419 & -0.5925(4) & N/A & -0.5872(1) & -0.5930(2) \\ \hline
%2109 & -0.5931(6) & N/A & -0.5875(1) & -0.5938(1) \\ \hline
  \end{tabular}
  \caption{
The  range for the critical Wigner-Seitz radius, $r_s^c$, for the 14-, 54-, and 114-electron UEG models.
Above $r_s^c$, the R to U instability occurs and thus UHF solutions appear.
}
  \label{tab:rsc}%\footnote{test}
\end{table}
}

\subsection{The 14-Electron UEG Model}
We begin by studying the 14-electron UEG which was studied in detail by Shepherd and co-workers in Ref.~\citenum{shepherd_ueg_jcp}.
This small benchmark system is helpful as it is accessible to most quantum chemistry methods, whilst still exhibiting the typical challenges one faces when simulating real solids, namely basis set incompleteness error, and strong correlation (when $r_s$ is large). In addition, it has of late emerged as a standard benchmark system for the UEG\citep{neufeld_ccmc}. 

\subsubsection{Basis Set Convergence}
\begin{figure}
\includegraphics{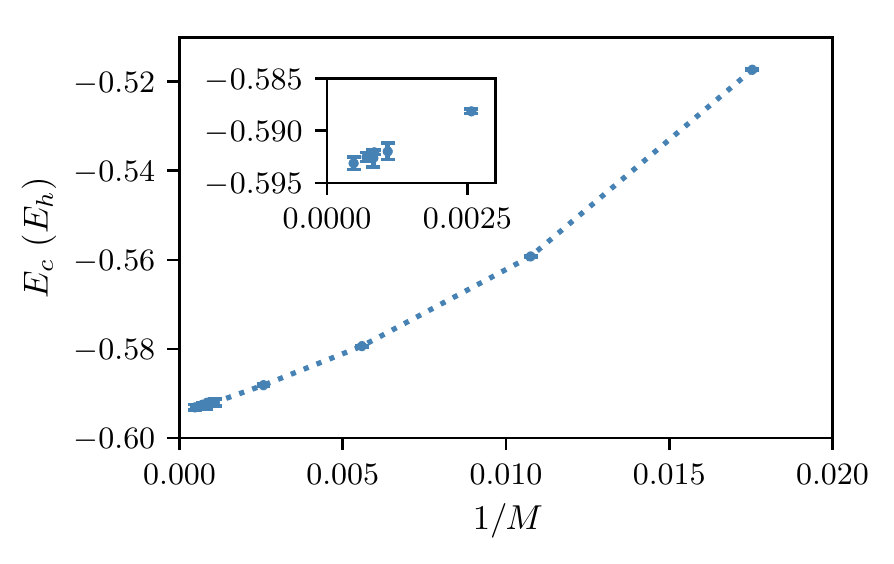}
\caption{
The basis set convergence in the ph-AFQMC correlation energy, $E_c$, of the 14-electron UEG model at $r_s=0.5$.
The basis sets considered are $M= 57, 93, 179, 389, 925, 1189, 1213, 1419, 2109$.
The ph-AFQMC values are reproduced in \cref{tab:rs05} for clarity.
The inset shows the correlation energy for $M\ge 389$.
Note that the onset of a clear $1/M$ dependence occurs at least past $M=925$.
}
\label{fig:basis}
\end{figure}

%The basis convergence is slower with lower $r_s$ values. 
%This is due to the fact that at lower $r_s$ the average distance between electrons is shorter.
%At shorter distances, being able to describe electron-electron cusps is crucial to obtain quantitatively accurate results.
%As it is typically the case in gaussian type orbitals (GTOs), resolving electron-electron cusps also require
%a lot of PWs.
%It is sufficient to check the basis set convergence with $r_s=0.5$ (the smallest $r_s$ value) 
%and we can assume higher $r_s$ values would converge faster to the CBS limit.
The basis set convergence of wavefunction based quantum chemistry methods for the UEG has been explored a number of times by various methods\citep{shephed_ueg,shepherd_wfn_conv,shepherd_mbqchem} and we will only briefly comment on it here. 
For our purposes, it is sufficient to note that the convergence to the CBS limit is slowest at high densities (low $r_s$) and thus it is sufficient to converge the basis set error here.
This slower convergence can be understood simply because the electron-electron cusp is more pronounced at high densities (the electron are more likely to coalesce).
This is seen in \cref{fig:basis} and \cref{tab:rs05} where on the order of 2000 PWs are necessary to converge the total energy to within 1 mHa in absolute energy (not per electron). 
Similar to previous studies we observe a more or less linear relationship between $E_c$ and $1/M$ for $M$ greater than 925\citep{shephed_ueg,neufeld_ccmc}.
\insertnew{The linear extrapolation to the CBS limit when using PWs was thoroughly studied and understood by Shepherd and
co-workers in Ref. \citenum{shepherd_wfn_conv}.}
We point out that the use of transcorrelated approaches in AFQMC could greatly accelerate the convergence to the CBS limit\citep{alavi_trans}.

\begin{table}
  \centering
  \begin{tabular}{r|r|r|r|r}\hline
$M$ & ph-AFQMC & $i$-FCIQMC & \multicolumn{1}{c|}{RCCSD} & \multicolumn{1}{c}{RCCSDT} \\ \Xhline{3\arrayrulewidth}
57 & -0.5173(1) & -0.5169(1) & N/A & N/A \\ \hline
93 & -0.5592(2) & -0.5589(1) & N/A & N/A \\ \hline
179 & -0.5794(2) & -0.5797(3) & -0.57365(1) & -0.57971(3) \\ \hline
389 & -0.5881(2) & -0.5893(3) & N/A & N/A \\ \hline
925 & -0.5920(8) & -0.5936(3) & -0.58626(4) & -0.5923(1) \\ \hline
1189 & -0.5921(2) & -0.5939(4) & N/A & N/A \\ \hline
1213 & -0.5926(8) & N/A & N/A & N/A \\ \hline
1419 & -0.5925(4) & N/A & -0.5872(1) & -0.5930(2) \\ \hline
2109 & -0.5931(6) & N/A & -0.5875(1) & -0.5938(1) \\ \hline
  \end{tabular}
  \caption{
  The correlation energy comparison between ph-AFQMC, $i$-FCIQMC, RCCSD, and RCCSDT for the 14-electron UEG model at $r_s=0.5$.
The $i$-FCIQMC numbers were taken from Ref.~\citenum{shephed_ueg} and CC numbers were taken from Ref.~\citenum{neufeld_ccmc}.
  N/A means that the data is not available. These calculations were performed using the \texttt{PAUXY} package. Error bars were estimated using reblocking\citep{flybjerg_blocking} as implemented in the \texttt{pyblock} package\citep{pyblock}.
}
  \label{tab:rs05}%\footnote{test}
\end{table}

FCIQMC with the initiator approximation ($i$-FCIQMC) is a formally exact approach as long as there is no
initiator bias. Therefore, comparing ph-AFQMC with $i$-FCIQMC is a good way to assess the accuracy of ph-AFQMC.
We see that ph-AFQMC agrees well with $i$-FCIQMC within the error bar up to $M=179$ and it starts to deviate from $i$-FCIQMC beyond that.
However, $i$-FCIQMC numbers for the larger $M$ values should be taken cautiously as it has been noted elsewhere that 
the bias from the initiator approximation was not completely removed\citep{neufeld_ccmc} and that the $i$-FCIQMC results for $r_s=0.5$ may be too low by approximately 1 m$E_h$.

Comparing ph-AFQMC and CC methods is perhaps more relevant for the purpose of this paper.
The UEG model $r_s=0.5$ is relatively weakly correlated and thus CC methods on top of an RHF reference work very well.
Neufeld and Thom found that for $r_s=0.5$, CCSDT is enough to converge the correlation energy with respect to the excitation levels\citep{neufeld_ccmc}.
Therefore, the CCSDT numbers in Table \ref{tab:rs05} should be considered to be exact for a given basis set.
As it is clear from Table \ref{tab:rs05},
the CCSD correlation energies are all above those of ph-AFQMC and ph-AFQMC agrees with CCSDT up to sub millihartree.

These results are particularly encouraging for the following reasons.
This dense UEG model may be analogous to a weakly correlated molecular system, in the sense that for a finite number of electrons it is relatively well described by HF theory.
In such a system, CCSDT (or CCSD(T)) should be more or less exact. Even if their absolute energies were not exact, the relative energies
such as barrier heights and interaction energies should be close to exact. 
The results here suggest that ph-AFQMC is a potentially powerful tool to handle such weakly correlated systems.
For the rest of this section, 
we will assess the accuracy of ph-AFQMC for higher $r_s$ where
there can be a good mixture of weak and strong correlation.
Furthermore, the quality of the RHF trial wavefunction will start to degrade so we will show how this affects ph-AFQMC.

%\onecolumngrid 

\begin{table}
%  \centering
%  \begin{figure}
%  \includegraphics[scale=0.5]{table2.pdf}
%  \caption{
%The correlation energy comparison between ph-AFQMC, $i$-FCIQMC, RCCSD, and RCCSDT for the 14-electron UEG model at $r_s=1.0, 2.0, 3.0$ and $5.0$.
%The $i$-FCIQMC numbers were taken from Ref. \citenum{shepherd_ueg_jcp} and CC numbers were taken from Ref. \citenum{neufeld_ccmc}.
%  N/A means that the data is not available.
%}
%  \label{tab:14e}
%  \end{figure}
  \begin{tabular}{c|r|r|r|r|r}\hline
$M$ & \multicolumn{1}{c|}{ph-AFQMC} & \multicolumn{1}{c|}{$i$-FCIQMC} & \multicolumn{1}{c|}{RCCSD} & \multicolumn{1}{c|}{RCCSDT} & \multicolumn{1}{c}{RCCSDTQ}\\ \Xhline{3\arrayrulewidth}
%%%%%%%%%%%%%%%%%%%%%%%
\multicolumn{6}{c}{$r_s$=1.0} \\ \hline
179 & -0.5187(6) & N/A & -0.50250(7) & -0.51819(3) & -0.51856(7) \\\hline
1189 & -0.5302(3) & -0.5305(5) & N/A & N/A & N/A \\\hline
2109 & -0.5298(8) & N/A & -0.5133(3) & -0.5290(4) & N/A \\\Xhline{3\arrayrulewidth}
%%%%%%%%%%%%%%%%%%%%%%%
\multicolumn{6}{c}{$r_s$=2.0} \\ \hline
81 & -0.4181(2) & N/A & -0.40181(4) & -0.41339(3) & -0.41579(2)\\\hline
925 & -0.4438(6) & -0.4431(5) & -0.4077(1) & -0.4388(1) & N/A \\ \hline
2109 & -0.4420(9) & N/A & -0.4089(2) & N/A & N/A \\ \Xhline{3\arrayrulewidth}
\multicolumn{6}{c}{$r_s$=3.0} \\ \hline
81 & -0.3590(2) & N/A & -0.32208(3) & -0.35671(3) & -0.36141(5)\\\hline
179 & -0.3723(4) & N/A & -0.3347(1) & -0.37246(5) & N/A\\\hline
925 & -0.3725(5) & N/A & -0.3389(2) & N/A & N/A\\\Xhline{3\arrayrulewidth}
%2109 & -0.3770(10) & N/A & -0.3392(3) & N/A & N/A \\\Xhline{3\arrayrulewidth}
\multicolumn{6}{c}{$r_s$=5.0} \\ \hline
179 & -0.2701(1) & -0.3017(7) & -0.2510(1) & -0.2925(1) & N/A \\\Xhline{3\arrayrulewidth}
  \end{tabular}
  \caption{
The correlation energy comparison between ph-AFQMC, $i$-FCIQMC, RCCSD, and RCCSDT for the 14-electron UEG model at $r_s=1.0, 2.0, 3.0$ and $5.0$.
The $i$-FCIQMC numbers were taken from Ref. \citenum{shepherd_ueg_jcp} and CC numbers were taken from Ref. \citenum{neufeld_ccmc}.
  N/A means that the data is not available.
}
  \label{tab:14e}
\end{table}
\twocolumngrid 

\subsubsection{Assessment for lower densities}

In Table \ref{tab:14e}, we present the comparison of ph-AFQMC, $i$-FCIQMC, RCCSD, RCCSDT, and RCCSDTQ for selected basis sets.
All of our ph-AFQMC is available in the Supplementary Materials.
At $r_s=1.0$, ph-AFQMC agrees with $i$-FCIQMC within the error bar of each result when $M = 1189$.
Small basis set ($M=179$) results suggest that from RCCSDT to RCCSDTQ only small correlation energy is gained. Therefore, we consider RCCSDT to be near-exact for larger basis sets.
Near the CBS limit ($M=2109$), we found that the ph-AFQMC energy is 15-16 m$E_h$ lower than RCCSD and is within the error bar of RCCSDT.

At $r_s = 2.0$ ph-AFQMC agrees with $i$-FCIQMC within each error bar. However, RCCSDT struggles to obtain quantitatively accurate results for $M=925$.
RCCSD is about 36 m$E_h$ above and RCCSDT is about 5 m$E_h$ above the $i$-FCIQMC (and ph-AFQMC) correlation energies.
Like usual strongly correlated systems, the effect of quadruples is not negligible here and it accounts for about 2 m$E_h$ correlation energy in a small basis ($M=81$).
As shown in \cref{tab:14e}, ph-AFQMC provides a lower correlation energy than even RCCSDTQ in the $M=81$ basis set.
\insertnew{Since neither ph-AFQMC nor RCCSDTQ is variational, such correlation energy comparisons should be taken with caution.
Nevertheless, since ph-AFQMC and $i$-FCIQMC agree for $M=925$ we expect ph-AFQMC to be accurate (i.e., near-exact) for $M=81$.}
This result highlights the utility of RHF+ph-AFQMC. Namely, it can provide quantitatively accurate results when the role of quadruples is not negligible and yet still small enough for the RHF trial wavefunction to behave well.

Although at $r_s =0.5, 1.0, 2.0$ ph-AFQMC provides more or less exact correlation energies, as the density is lowered further we find that the RHF trial wavefunctions performance degrades significantly.
Not only does the ph-AFQMC correlation energy become above RCCSDTQ by about 2 m$E_h$ at $r_s=3.0$, but the stability of the simulations suffers noticeably.
Nevertheless, at $r_s = 3.0$ we observe that ph-AFQMC is comparable to RCCSDT and is able to reach the CBS limit reliably.

\begin{figure}
\includegraphics{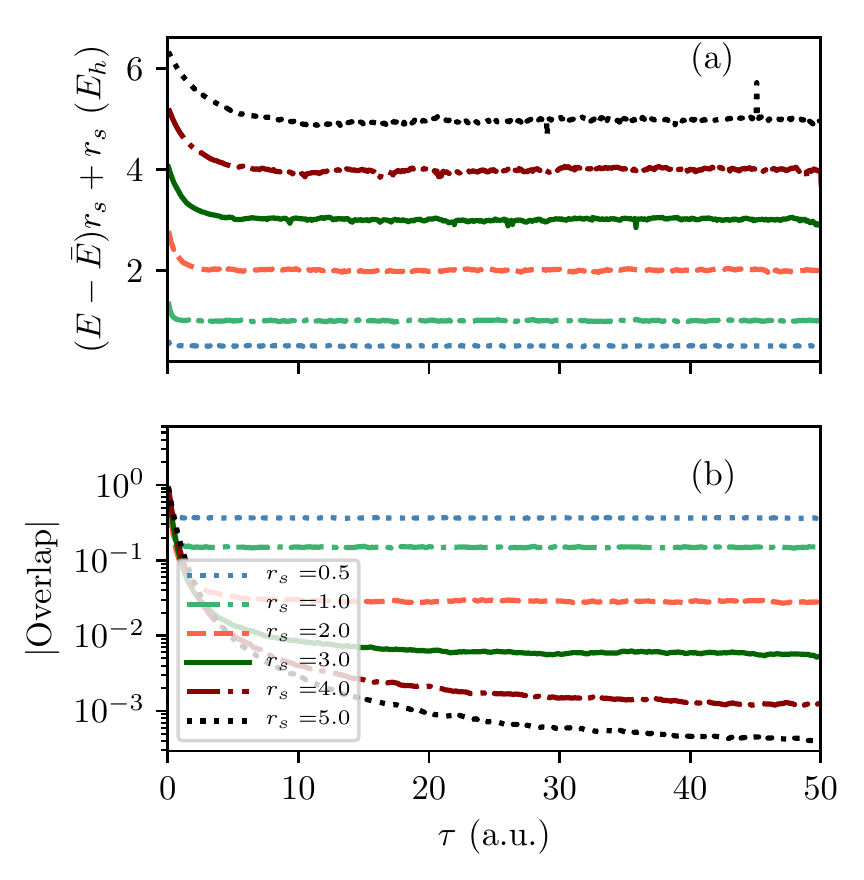}
\caption{Panel (a) shows the convergence of the ph-AFQMC total energy to its equilibrated mean value ($\bar{E}$) as a function of projection time for a variety of values of $r_s$. Note that the data have been shifted and scaled by $r_s$ for clarity. Note the occurrence of  spikes in the local energy increases with $r_s$. Here we bounded the local energy during propagation but not when printing the estimator to reveal the degradation in the results. Panel (b) plots the reduction in the magnitude in the overlap between the walkers and the trial wavefunction (see main text for definition) with decreasing $r_s$. The slower equilibration of the overlap compared to the local energy has been noted previously in Ref.~\citenum{purwanto_excited_c2}. \label{fig:overlap}}
\end{figure}

However, $r_s=5.0$ is much more difficult to handle with an RHF trial wavefunction. This is typically evidenced by an increase in the number of rare event population fluctuation.
These rare events are well understood, and arise due a divergent importance function which occurs when $\langle \Psi_T|\phi\rangle$ approaches zero.
Although these rare events can be effectively controlled by the use of bounds on the local (and/or hybrid energy)\citep{purwanto_pressure_bound} they nevertheless signify a worsening in the quality of trial wavefunction for a fixed system size.
To demonstrate this, in \cref{fig:overlap} we plot the convergence of the ph-AFQMC energy with projection time for a range of densities with $M=93$ as well as an estimate for the overlap $\sum_n w_n |\langle \Psi_T|\phi_n\rangle|/\sum_n w_n$.
We see from \cref{fig:overlap} (a) that as $r_s$ increases the projection time necessary to converge to the ground state increases as well as the frequency of rare events.
This is correlated with a decrease in the magnitude in the overlap as is seen from \cref{fig:overlap} (b).

Indeed, at $r_s=5.0$ we found that the ph-AFQMC energy did not converge monotonically with increasing basis set size past $M=389$. 
Rather, the ph-AFQMC correlation energy decreases in magnitude with increasing  basis set size.
This signals a complete breakdown of the phaseless constraint with this trial wavefunction.
We note that a similar effect can be observed in $i$-FCIQMC when the initiator error is not fully converged for increased basis set sizes, where one finds that the correlation energy begins to plateau as a function of basis set.
This suggests that an improved trial wavefunction is necessary to attain sensible results for this system.

It is noteworthy to point out that this unusual behavior of ph-AFQMC energy with respect to the basis set size could indicate the ``non-variational'' failure of ph-AFQMC. ph-AFQMC is formally non-variational
in the sense that a variational energy estimator of a given ph-AFQMC wavefunction can be above the mixed energy estimator in Eq. \eqref{eq:mixedE}.\cite{carlson_no_var}
Similarly, CC methods are also formally non-variational due to their projective nature. With an RHF reference, it has shown catastrophic non-variationality
for strongly correlated systems.\cite{van2000benchmark,cooper2010benchmark,lee2017coupled} It is possible that RCCSD (and even RCCSDT) is also exhibiting non-variationality for this $r_s$ value.
This can be confirmed with more sophisticated CC methods.\cite{van2000quadratic,robinson2011approximate}
The investigation of the non-variationality of ph-AFQMC and CC methods in the context of strong correlation will be an interesting subject for future study.

\insertnew{
As discussed in \cref{subsec:uhf}, above $r_s=3.5$, UHF solutions appear
and they can be often powerful trial wavefunctions for ph-AFQMC \cite{purwanto2008eliminating}.
As in Ref. \citenum{purwanto2008eliminating}, 
we employ the spin-projection technique to remove the spin-contamination completely.
That is, 
we use 
the RHF wavefunction as the initial wavefunction
while using the UHF wavefunction for the constraints.
As shown in \cref{sec:size},
because both RHF and UHF are size-consistent
the resulting UHF+ph-AFQMC performed with an RHF initial wavefunction must be also size-consistent.
In \cref{tab:rs5_msd}, we present the correlation energies from UHF+ph-AFQMC.
Given its substantial symmetry breaking at $r_s=5.0$ shown in \cref{fig:uhf},
it is surprising that it does not provide much improvement over RHF+ph-AFQMC.
The UHF+ph-AFQMC energies are about 1 mH lower than the RHF+ph-AFQMC energies and they
are still far away from more accurate $i$-FCIQMC benchmark energies.
}

\begin{figure}
\includegraphics{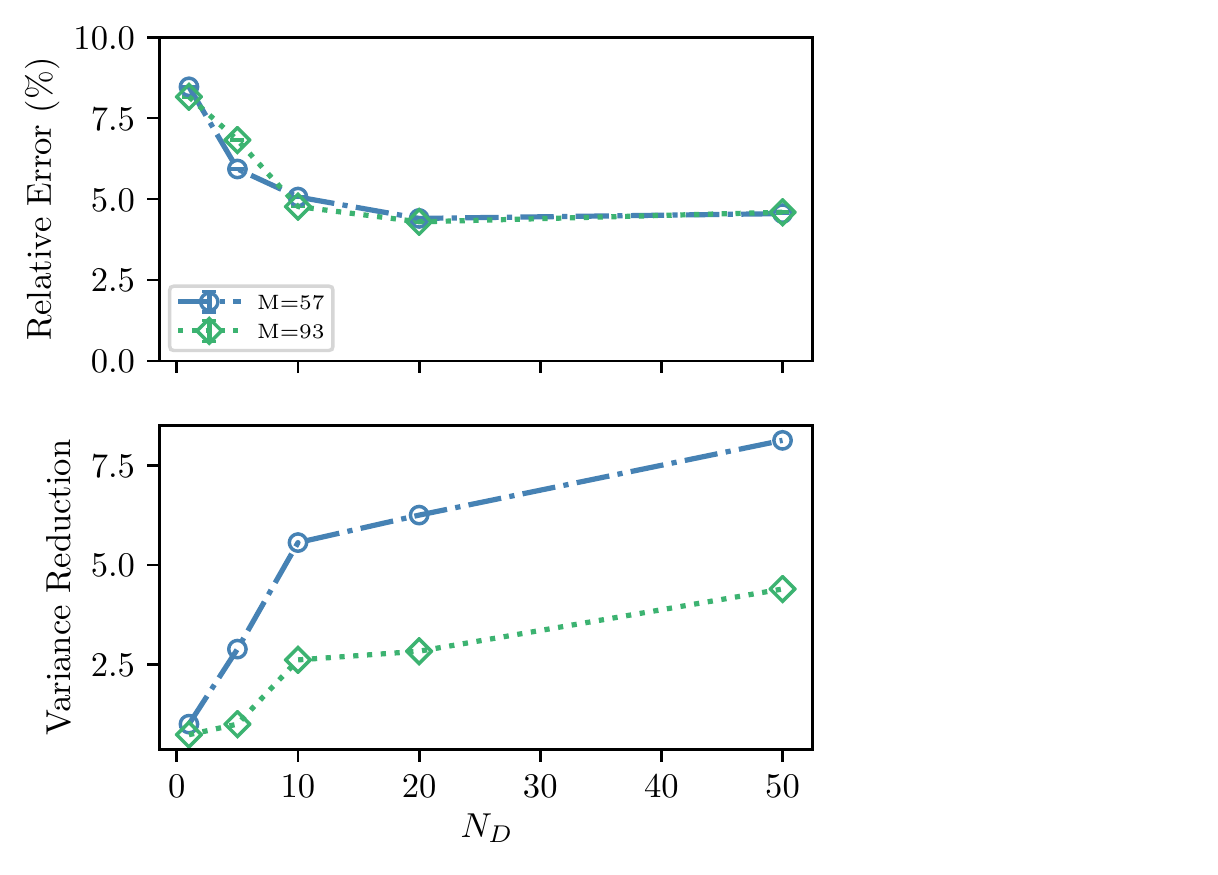}
\caption{Panel (a) shows the behavior of the relative error in the ph-AFQMC correlation energy as a function of the number of determinants in the trial wavefunction expansion, $N_D$. The relative error is measured with respect to the $i$-FCIQMC value. Panel (b) shows the corresponding reduction in statistical variance defined as $\mathrm{Var}(N_D=1)/\mathrm{Var}(N_D)$.\label{fig:msd}}
\label{fig:var}
\end{figure}

To investigate the ph-AFQMC results at $r_s=5.0$ further, we explored the use of non-orthogonal multi-Slater determinant expansions (NOMSD) as trial wavefunctions generated using a version of
the projected Hartree–Fock (PHF) algorithm developed by Scuseria and co-workers\citep{Jimenez_phf,scuseria_phf,scuseria_phf_grad}. Interested readers are referred to Ref.\citenum{borda2018non} for further details.
In \cref{fig:var}, We find an initial rapid decrease in the error of the ph-AFQMC correlation energy and correspondingly a reduction in the ph-AFQMC statistical variance in the local energy estimator.
\insertnew{
The statistical variance in the estimator
should not be confused with the wavefunction variance in variational MC.
Therefore, it is possible to observe 
some improvement in the statistical error bar while 
the energy estimator does not improve noticeably as observed before\cite{borda2018non}.
}
This long tail in the convergence of the ph-AFQMC energy is indicative that the system is strongly correlated.
We note that the FCI space for $M=57$ contains on the order of $10^{16}$ determinants and therefore the improvement in the trial wavefunction via determinantal expansions is eventually limited by the exponential wall.
We find similar behavior for larger basis sets. 

\cref{tab:rs5_msd} summarizes our ph-AFQMC results using a 10 determinant expansion.
Note that the $M=93$ and $M=389$ values are roughly within error bars of each other, despite the $i$-FCIQMC correlation energy decreasing by approximately 10 m$E_h$. We found that for even larger basis sets the RHF+ph-AFQMC correlation energies lay \emph{above} those at $M=93$.
Nevertheless we see that by improving the trial  wavefunction the ph-AFQMC correlation energies begin to slowly approach those of $i$-FCIQMC values. The slow convergence is evidence of the limitation of using multi Slater determinant trial wavefunctions in strongly correlated systems. 

\begin{table}
  \centering
  \begin{tabular}{r|r|r|r|r}\hline
$M$ & RHF & UHF & NOMSD &  $i$-FCIQMC
\\ \hline
57 & -0.2422(8) &-0.24371(9)& -0.2511(3) & -0.2645(3)\\ \hline
93 & -0.2677(5) & -0.26837(8)& -0.2786(2) & -0.2928(4) \\ \hline
389 & -0.2674(6) & -0.2654(2) & -0.2794(4) & -0.304(1) \\
\hline
\end{tabular}
\caption{Comparison between ph-AFQMC correlation energies using a RHF\insertnew{, a UHF} and ten determinant NOMSD trial wavefunction at $r_s=5.0$. Correlation energies are measured relative to the RHF total energy. $i$-FCIQMC energies were taken from Ref.~\citenum{shepherd_ueg_jcp}. ph-AFQMC calculations were performed using the development version of QMCPACK.
Note that the $M=389$ RHF+ph-AFQMC energy is above the $M=179$ RHF+ph-AFQMC energy in \cref{tab:14e} and this is an artifact of the breakdown of the RHF trial wavefunction.
%\insertnew{Note that the UHF+AFQMC correlation energy was measured
%with respect to the RHF energy.}
\label{tab:rs5_msd}
}
\end{table}

\subsection {Larger Supercells}
As explained in Section \ref{sec:size},
ph-AFQMC is size-consistent and thus can reliably reach the thermodynamic limit
using larger super cells
along combined with finite size corrections and twist averaging\citep{ChiesaRPA06,KZK2008,KZK2011,drummond_fzc,holzmann_fzc,lin_twist_av}.
As the total energy of the UEG model in the thermodynamic limit is already well understood for solid state densities\citep{ceperley_alder}
here we instead just study finite-sized UEG models and compare with other available methods when applicable.

Following the sequence of ``magic numbers'' in the UEG model,
we study larger supercells (54 electrons and 114 electrons) with ph-AFQMC for $r_s=0.5, 1.0, 2.0$.
In the 14-electron UEG model, we obtained energies with error bars of the order of 1 m$E_h$. 
This is important for molecular applications where we aim for energy differences between two finite systems.
On the other hand, the cost of achieving the same statistical error for larger systems adds an extra $\mathcal O(N)$ 
to the computational cost of ph-AFQMC.
This extra cost for sampling may be avoided by the correlated sampling technique\citep{shee_chem_trans,shee_gpu}, but here we instead
compare the total energy {\it per electron}.
This metric is well-suited for ab-initio solids (or extended systems) in general.

High-quality DMC numbers are available for the 54-electron UEG model
and we compare ph-AFQMC against this.
For the 114-electron UEG model, there are only variational MC (VMC) results available
so we will compare against these.

\subsubsection{The 54-Electron UEG Model}
We found that from $M = 1419$ to $M = 2109$ the change in $E_\text{tot}/N$ at $r_s=0.5$ is only of the order of 0.1 m$E_h$.
We therefore do not perform the CBS extrapolation for the comparison between ph-AFQMC and DMC.
The reported ph-AFQMC numbers are obtained from $M=1419$ which enables a direct comparison between
ph-AFQMC and $i$-FCIQMC at $r_s = 0.5$ and $r_s=1.0$.
The ph-AFQMC results for $M=2109$ at $r_s=0.5, 1.0, 2.0$ are available in the Supplementary Materials.

\begin{table}
  \centering
  \begin{tabular}{c|r|r|r|r}\hline
$r_s$ & \multicolumn{1}{c|}{ph-AFQMC} & \multicolumn{1}{c|}{$i$-FCIQMC} 
& \multicolumn{1}{c|}{FN-DMC-SJ} & 
\multicolumn{1}{c}{FN-DMC-BF}\\ \hline
0.5 & 3.22087(2) & 3.22086(2) & 3.22245(9) & 3.22112(4)\\ \hline
1 & 0.52967(2) & 0.53073(4) & 0.53089(9) & 0.52989(4)\\ \hline
2 & -0.01429(3) & N/A & -0.01311(2) & -0.013966(9)\\ \hline
5 & -0.07589(5) & N/A & -0.078649(7) & -0.079036(3)\\ \hline
\end{tabular}
  \caption{
The total energy per electron ($E_h$/e) comparison between ph-AFQMC, $i$-FCIQMC, FN-DMC with a Slater-Jastrow (SJ) trial wavefunction, and DMC with a backflow (BF) trial wavefunction for the 54-electron UEG model at $r_s=0.5,1.0, 2.0, 5.0$.
Both ph-AFQMC and $i$-FCIQMC numbers are obtained from $M=1419$.
The $i$-FCIQMC numbers were taken from Ref.~\citenum{shephed_ueg} and DMC numbers were taken from Ref.~\citenum{rios2006inhomogeneous}.
N/A means that the data is not available.
}
  \label{tab:54e}
\end{table}

In Table \ref{tab:54e}, we summarize the comparison between ph-AFQMC, $i$-FCIQMC, and DMC for the 54-electron UEG model.
At $r_s=0.5$, ph-AFQMC and $i$-FCIQMC agree with each other within the error bar. 
Shepherd and co-workers found that the DMC-BF energy is somewhat higher than $i$-FCIQMC and suggested that the fixed-node error with the backflow (BF) trial wavefunction may not be small.
Indeed, we reach the same conclusion with ph-AFQMC.
As explained in \cref{sec:intro}, the difference between DMC and AFQMC is mainly the discretization (or basis set) we work with.
It is interesting that the fixed-node error in FN-DMC can be non-negligible even with more sophisticated trial wavefunctions 
such as Slater-Jastrow (SJ) and BF.
It is encouraging that we can achieve near-exact accuracy with ph-AFQMC at $r_s=0.5$ with this simplest possible RHF trial wavefunction.

\revrep{At $r_s=1.0$, we observe that ph-AFQMC performs better than $i$-FCIQMC and is in agreement with FN-DMC-BF up to 0.2 m$E_h$/e.}{At $r_s=1.0$, ph-AFQMC is in a better agreement (the difference is about 0.2 m$E_h$/e) with FN-DMC-BF than $i$-FCIQMC is.} \insertnew{In this case, $i$-FCIQMC suffers from the initiator bias and results into about 1 m$E_h$/e above the FN-DMC-BF energy.} In fact, the ph-AFQMC energy is lower than that of FN-DMC-BF, which may indicate non-negligible fixed-node errors even in FN-DMC-BF. 
\insertnew{Since ph-AFQMC is not variational while FN-DMC-BF is, more careful calibration is highly desirable
to see if there are indeed fixed-node errors in FN-DMC-BF.}\revrep{In the $M=2109$ basis set, the difference becomes 0.3 m$E_h$/e as shown in the Supplementary Materials.}{}\revrep{In this case, $i$-FCIQMC suffers from the initiator bias and results into about 1 m$E_h$/e above the FN-DMC-BF energy.}{}
\revrep{The performance of ph-AFQMC is better than FN-DMC-SJ by 1 m$E_h$/e similarly to the $r_s=0.5$ case.}
{ph-AFQMC agrees better with FN-DMC-BF than does FN-DMC-SJ by 1 m$E_h$/e similarly to the $r_s=0.5$ case.}

No $i$-FCIQMC results at $r_s = 2.0$ due to the severity of the sign problem so we instead must compare only to FN-DMC.
We find that the ph-AFQMC energy is 1.1 m$E_h$/e below the FN-DMC-SJ energy and 0.3 m$E_h$/e below the FN-DMC-BF energy. With a larger basis set $M = 2109$, the ph-AFQMC energy lies 0.4 m$E_h$/e below the FN-DMC-BF energy as shown in the Supplementary Materials.
Further increasing $r_s$ to 5.0, we observe that ph-AFQMC is no longer comparable to FN-DMC-SJ and FN-DMC-BF as expected. \revrep{It should be noted that despite the severe sign problem at $r_s =5.0$ ph-AFQMC is still able to provide an answer with good statistics owing to the phaseless approximation (though the correlation energy error is significant).}{}\insertnew{As the use of UHF and NOMSD trials was found to be ineffective in the 14-electron model at $r_s=5.0$,
we did not perform such calculations here.}

In summary, for the 54-electron UEG model at $r_s = 0.5, 1.0, 2.0$, we observe that ph-AFQMC can obtain nearly exact $E_\text{tot}/N$.
In particular, its accuracy is comparable to other state-of-the-art methods such as $i$-FCIQMC and FN-DMC-BF.
The general conclusions are similar to the 14-electron UEG model: ph-AFQMC is particularly well-suited for $r_s$ values smaller than 5 where there exists moderate strong correlation.
It is encouraging that ph-AFQMC achieved these highly accurate results using the simplest trial wavefunction, RHF.
\subsubsection{The 114-Electron UEG Model}
Encouraged by the near-exact accuracy of ph-AFQMC for low $r_s$ values in the 14- and 54-electron UEG models, we used ph-AFQMC to provide benchmark numbers for the 114-electron UEG model for future method development.
The 114-electron UEG model has been relatively less explored.
For determinant-based algorithms like $i$-FCIQMC the sign problem is likely to preclude its application except for very high densities.
On the other hand, for ph-AFQMC this does not pose a significant challenge especially when considering $r_s \le 2.0$.

At $r_s=0.5$, the total energy per electron changes by 0.5 m$E_h$/e when increasing $M$ from 1419 to 2109.
For higher $r_s$ values, we expect this energy change to be smaller.
We will present the ph-AFQMC energies at $r_s=0.5, 1.0, 2.0$ 
all obtained with $M=2109$.
%and
%the $r_s=0.5$ result was obtained with $M=2109$ and the other values of $r_s$ were obtained with $M=1419$.
We expect that our ph-AFQMC energies reported here have the basis set incompleteness error
of the order of 0.5 m$E_h$/e per electron. Therefore, the numbers reported here may be considered as an upper bound to the ph-AFQMC energies at the CBS limit.
%We present the ph-AFQMC energies in Table \ref{tab:114e}.
%We expect that the ph-AFQMC energies reported here are good up to 0.1

\begin{table}
  \centering
  \begin{tabular}{c|r}\hline
$r_s$ & \multicolumn{1}{c}{ph-AFQMC}\\\hline
0.5 & 3.48453(8) \\\hline
1.0 & 0.59877(6) \\\hline
2.0 & 0.00487(6) \\\hline
\end{tabular}
  \caption{
The total energy per electron ($E_h$/e) of ph-AFQMC for the 114-electron UEG model at $r_s=0.5,1.0, 2.0$.
All results were obtained with $M=2109$. %whereas the other results were obtained with $M=1419$.
The VMC (Slater-Jastrow) energy at $r_s=1.0$ is 0.60395(25) $E_h$/e.\cite{kwon1998effects}
%The $i$-FCIQMC numbers were taken from Ref. [] and DMC numbers were taken from Ref. [].
%N/A means that the data is not available.
}
  \label{tab:114e}
\end{table}

The ph-AFQMC results are presented in Table \ref{tab:114e}. The only data available in literature
is $r_s=1.0$ with a VMC approach with a Slater-Jastrow wavefunction.
The VMC energy is 0.60395(25) $E_h$/e \cite{kwon1998effects} which is at least 5 m$E_h$/e higher than our ph-AFQMC energies.
Comparing ph-AFQMC energies in Table \ref{tab:54e} and Table \ref{tab:114e},
we note that the finite-size effect is still very large.
Namely, the energy per electron is far from the convergence with respect to the system size.
It will be interesting to investigate finite-size effects with ph-AFQMC in more realistic systems in the future.
Although further comparisons are not possible due to the lack of benchmark data,
we believe that the ph-AFQMC numbers in Table \ref{tab:114e}
are close to the exact energies and the correlation energy error is smaller than 1 m$E_h$ {\it per electron} given the results for the 54-electron model.

\section{Conclusions}
In this paper, we examined the performance of phaseless auxiliary-field quantum Monte Carlo (ph-AFQMC)
with the spin-restricted Hartree-Fock (RHF) trial wavefunction (i.e., RHF+ph-AFQMC)
on the uniform electron gas (UEG) problem.
We considered the 14-electron, 54-electron, and 114-electron UEG model.
Through these studies, we found the following conclusions:
\begin{enumerate}
\item In the 14-electron case, we compared RHF+ph-AFQMC with spin-restricted coupled-cluster (RCC) methods.
Compared to RCC with singles and doubles (RCCSD) and CC with singles, doubles, and triples (RCCSDT), 
RHF+ph-AFQMC performs better than RCCSDT and similarly to or slightly worse than RCCSDTQ for $r_s \le 3.0$.
\item 
For the 14-electron problem at $r_s = 5.0$ where CCSDT is inadequate, RHF+ph-AFQMC exhibits rare fluctuations in the energy estimator,
which makes the phaseless approximation difficult to use.
We found that using a small multi-determinant trial wavefunction is effective in stabilizing the simulations but still ineffective in obtaining highly accurate results in such cases.
\item In the case of the 54-electron UEG model,
the comparison with initiator full configuration interaction QMC ($i$-FCIQMC) and fixed-node diffusion MC (FN-DMC)
suggested that RHF+ph-AFQMC is a promising tool for simulating dense solids. 
\insertnew{Such connections between dense solids and the UEG model were previously made in Ref. \citenum{salpeter1961energy, baus1980statistical, huotari2010momentum}.}
RHF+ph-AFQMC confirmed that the fixed-node error in FN-DMC for $r_s=0.5$ as noted before in an $i$-FCIQMC study.
Moreover, RHF+ph-AFQMC revealed that the initiator bias in $i$-FCIQMC for $r_s=1.0$ is large (about 1 m$E_h$ per electron) and the fixed-node error in FN-DMC with a back flow trial wavefunction (FN-DMC-BF) may not be negligible (0.3 m$E_h$ per electron). A smilar trend was observed in the case of $r_s=2.0$. Lastly, $r_s=5.0$ was found to be challenging for RHF+ph-AFQMC to tackle and the ph-AFQMC correlation energy was simply inadequate compared to FN-DMC-BF for this case.
Overall, RHF+ph-AFQMC was found to be as accurate as or potentially more accurate than FN-DMC-BF wavefunction for $r_s$ up to 2.0.
\item
We produced RHF+ph-AFQMC energies of the 114-electron problem for $r_s \le 2.0$ where not many benchmark data are available.
Given its performance for the 54-electron case, 
we expect the RHF+ph-AFQMC correlation energy error to be less than 1 m$E_h$ per electron.
\end{enumerate}

It is the central message of this paper that even with the simplest trial wavefunction (RHF)
ph-AFQMC is a powerful tool for simulating molecules and solids where there is no noticeable strong correlation between electrons.
In particular, its scope lies between CCSD and CCSDT.
Given its low scaling ($\mathcal O(N^3) - \mathcal O (N^4)$), RHF+ph-AFQMC remains a promising tool.

The future study should include a more extensive benchmark of RHF+ph-AFQMC on more chemically relevant systems such as the W4-11\citep{karton_w411} set
as well as designing better and yet compact trial wavefunctions for AFQMC. Using dynamically correlated orbitals such as those from orbital-optimized M{\o}ller-Plesset perturbation theory can be an economical way to go beyond HF trial wavefunctions.\cite{lee2018regularized,lee2019distinguishing,lee2019two}
Some essential symmetry breaking in the HF trial wavefunction can potentially improve the performance of ph-AFQMC greatly such as using complex, restricted HF orbitals.\cite{Small2015,lee_dft}
Lastly, the finite-temperature extension of ph-AFQMC has been well established\citep{zhang_finite_t_prl,rubenstein_bose_fermi,liu_ft_afqmc}. The assessment of ph-AFQMC for the warm dense UEG model\citep{dornheim_wdueg_rev}, which has been the subject of intense research of late\citep{BrownUEG1,schoof_prl,malone_accurate_2016,dornheim_qmc_review,dornheim_dynamical} is currently work in progress.

\section{Acknowledgement}
The work of J.L. was partly supported by the CCMS summer internship in 2018 at the Lawrence Livermore National Lab.
J.L. thanks Martin Head-Gordon and Soojin Lee for consistent encouragement.
We would like to thank Carlos Jimenez-Hoyos for providing us access to
the PHF code used to generate NOMSD expansions. The PHF
code used in this work was developed in the Scuseria group at
Rice University \citep{Jimenez_phf,scuseria_phf,scuseria_phf_grad}.
This work was performed under the auspices of the U.S. Department of Energy
(DOE) by LLNL under Contract No. DE-AC52-07NA27344.  Funding support was from
the U.S. DOE, Office of Science, Basic Energy Sciences, Materials Sciences and
Engineering Division, as part of the Computational Materials Sciences Program
and Center for Predictive Simulation of Functional Materials (CPSFM).  Computer
time was provided by the Livermore Computing Facilities.
%We thank Hao Shi for useful conversations on the FT-AFQMC algorithm.
\bibliography{refs}
\end{document}